\documentclass{aastex}
\usepackage{natbib}
\shorttitle{Compositional Diversity of Extrasolar Terrestrial Planets}
\shortauthors{Carter-Bond et al.}

\begin{document}

\title{The Compositional Diversity of Extrasolar Terrestrial Planets: II. Migration Simulations}

\author{Jade C. Carter-Bond\altaffilmark{1}}
\affil{School of Physics, University of New South Wales, Kensington, NSW 2052}
\email{j.bond@unsw.edu.au}

\and

\author{David P. O'Brien}
\affil{Planetary Science Institute, 1700 E. Fort Lowell, Tucson, AZ 85719}

\and

\author{Sean N. Raymond}
\affil{Universit\'{e} de Bordeaux, Observatoire Aquitain des Sciences de l'Univers, 2 rue de l'Observatoire, BP 89, 33271, Floirac Cedex, France.}
\affil{CNRS, UMR 5804, Laboratoire d'Astrophysique de Bordeaux, 2 rue de l'Observatoire, BP 89, 33271, Floirac Cedex, France}

\altaffiltext{1}{Formerly published as Jade C. Bond}

\begin{abstract}
Prior work has found that a variety of terrestrial planetary compositions are expected to occur within known extrasolar planetary systems. However, such studies ignored the effects of giant planet migration, which is thought to be very common in extra-solar systems. Here we present calculations of the compositions of terrestrial planets that formed in dynamical simulations incorporating varying degrees of giant planet migration.  We used chemical equilibrium models of the solid material present in the disks of five known planetary host stars: the Sun, GJ 777, HD4203, HD19994 and HD213240. Giant planet migration has a strong effect on the compositions of simulated terrestrial planets as the migration results large-scale mixing between terrestrial planet building blocks that condensed at a range of temperatures. This mixing acts to 1) increase the typical abundance of Mg-rich silicates in the terrestrial planets' feeding zones and thus increase the frequency of planets with Earth-like compositions compared with simulations with static giant planet orbits; and 2) drastically increase the efficiency of the delivery of hydrous phases (water and serpentine) to terrestrial planets and thus produce water worlds and/or wet Earths. Our results demonstrate that although a wide variety of terrestrial planet compositions can still be produced, planets with Earth-like compositions should be common within extrasolar planetary systems.
\end{abstract}


\section{Introduction}
Large-scale orbital migration of gas giant planets is generally thought to have occurred in many extrasolar planetary systems \citep{armitage:2007}, and perhaps even in our own Solar System \citep[e.g.][]{walsh:2011a}. The mechanism that is thought to cause the most dramatic migration takes place during the gaseous protoplanetary disk phase and is referred to as ``type II'' migration (\cite{papaloizou:1984,lin:1996}). Type II migration occurs when a planet becomes large enough to clear an annular gap in the disk, thus effectively coupling itself to the motion of the gas. As the gas viscously evolves, the majority of the mass flows inward onto the star and drags the planet inward with it. The planet thus migrates on the disk's viscous timescale of $\sim$10$^{5-6}$ years.  Migration ceases when the disk mass drops to a fraction of the planet mass.

As a giant planet migrates inward from where it formed via this mechanism, it encounters the building blocks of planets forming closer in.  During this migration, surprisingly little material impacts the planet directly.  Rather, small bodies in the giant planet's path are either gravitationally scattered onto exterior orbits or shepherded inward by mean motion resonances and follow the gas giant's inward migration \citep{mandell:2007,fogg:2005,fogg:2007,raymond:2006}.  The later orbital evolution of these small bodies is also influenced by gasdynamic effects such as gas drag.  Thus, a consequence of giant planet migration is to redistribute large amounts of solid material throughout the disk. The redistribution of solid material exerts great influence on both the final architecture of the planetary system and the composition of any terrestrial planets forming within the system. By shepherding large amounts of material throughout the system, giant planet migration directly impacts both the mass and location of the final terrestrial planets that form. Similarly, planetary feeding zones are no longer confined solely to material located immediately adjacent to the planet. Instead, planetary building blocks are sourced from throughout the disk, potentially producing a greater diversity of possible planetary compositions.

Numerous studies have examined the issue of terrestrial planet formation and survival during giant planet migration \citep[e.g.][]{walsh:2011b,raymond:2009,mandell:2007,gaidos:2007,raymond:2006,fogg:2009,fogg:2007,fogg:2005,armitage:2003}. However, few studies have considered the impact of migration on planetary composition. \cite{raymond:2006} and \cite{mandell:2007} estimated the water content of simulated planets based on a simple assumed initial distribution that was calibrated to match the water contents of asteroid types as inferred from meteorites, although the asteroid belt's structure may itself have been sculpted by Jupiter's migration \cite{walsh:2011a}.  No prior work has undertaken a detailed examination of other aspects of planetary composition and giant planet migration.

In our previous work \citep[e.g.][]{carter-bond:2012,bond:2010b} we showed that a diversity of terrestrial planet compositions is likely to exist in other planetary systems. These compositions largely reflect that of the planetary host star, with the stellar Mg/Si and C/O ratios exerting the strongest control over both the bulk mineralogy of the solid material present within the disk and hence the final compositions of the terrestrial planets. Under the assumption of chemical equilibrium, a C/O value below 0.8 implies that Si will be present within the system in the solid form as a silicate, primarily as either the SiO$_{4}$$^{4-}$ or SiO$_{2}$ building block. The precise form of this silicate is controlled by the Mg/Si value, ranging from pyroxene (MgSiO$_{3}$) (for Mg/Si$<$1) to olivine (Mg$_{2}$SiO$_{4}$) (for Mg/Si$>$2) with a combination of the two for 1$<$Mg/Si$<$2. If the C/O value is greater than 0.8, then carbide phases (C, SiC and TiC) will be present in large quantities in addition to these silicate species, resulting in C-rich planetary building blocks \citep{kuchner:2005,gaidos:2000}. A wide variation in these two key ratios has been observed for stars known to host giant planets \citep{petigura:2011,delgado-mena:2010} with predicted terrestrial planetary compositions ranging from relatively ``Earth-like" planets to those that are dominated by C \citep{carter-bond:2012,bond:2010b}.

To date, all such studies have neglected giant planet migration and have instead used as their inputs simulations of late-stage terrestrial accretion with the giant planets on fixed orbits. Thus, in these studies the planetary compositions primarily reflect the condensation sequence of the disk. As a result, systems with refractory-dominated (Ca, Al, Na) terrestrial planets located in the inner region, water-rich planets in the outer region and Mg-silicate planets in the intermediate region were produced \citep{bond:2010b}. Given the apparently common occurrence of giant planet migration within extrasolar planetary systems and the potential of giant planet migration to redistribute solid material within the system, it is necessary to revisit the issue and examine in detail the effects that giant planet migration may have on the bulk composition of the terrestrial planets in those systems.

In this study, we present detailed calculations of the bulk elemental composition of simulated terrestrial planets in systems with migrating giant planets. We primarily focus on terrestrial planets forming within 3 AU from the host star. This region was selected as it is the target region of current and future missions such as Kepler, is of considerable astrobiological interest as it spans the traditional habitable zone and is the region in which we are most likely to first be able to detect and characterize the composition of terrestrial planetary atmospheres, providing us with observationally-based limits on the composition. We consider three scenarios with giant planet migration and compare the terrestrial planets that form in each with a suite of in-situ simulations and our own prior studies. This study is the first of its kind to consider both giant planet migration and its effects on the composition of the terrestrial planet produced.  It is not only crucial in the development of a more self-consistent model of terrestrial planet formation, but also in placing more realistic constraints on the full range of possible terrestrial planet compositions within extrasolar planetary systems.

\section{Simulations}
\subsection{Dynamical Simulations}
In this study, we selected four unique dynamical situations in which to examine terrestrial planet formation: (1)Jupiter-mass planet located at 1AU and not migrating (hereafter referred to as in-situ); (2)Jupiter-mass body migrating from 5AU to 1AU (JD$_{5-1}$); (3)Jupiter-mass body migrating from 5AU to 0.25AU (JD$_{5-0.25}$); and (4)Jupiter-mass body migrating from 5AU to 0.25AU with an additional Saturn-mass body stationary at 9.5AU (JSD$_{5-0.25}$). The first two sets of simulations were run expressly for the purpose of this study, the JD$_{5-0.25}$ were first presented in \cite{raymond:2006}, and both the JD$_{5-0.25}$ and JSD$_{5-0.25}$ simulations were previously reported in \cite{mandell:2007} (as JD and JSD, respectively). While there are small differences in the initial conditions between the different sets of simulations (e.g. in the embryo masses and separations), the differences are relatively small, such that we can still make reasonable comparisons between different simulations \citep[e.g.][]{kokubo:2006}.  Each set of simulations is discussed below.

\textit{In-situ \& JD$_{5-1}$}: For both the in-situ and JD$_{5-1}$ situations, N-body simulations of terrestrial planet accretion were run using the SyMBA n-body integrator \citep{duncan:1998}. Following \cite{obrien:2006}, both Lunar- to Mars-mass embryos and a swarm of smaller planetesimals were included. Note that, as is common with these types of simulations, the planetesimals interact only with the embryos and not with each each other. In the case of the in-situ simulations, all embryos and planetesimals are distributed between 0.3 and 0.8AU with 1.64 M$_{\bigoplus}$ of solid material equally divided between planetesimals and embryos (M$_{embryos}$ = M$_{planetesimals}$ = 0.82 M$_{\bigoplus}$). The JD$_{5-1}$ situations began with a total of 12.19 M$_{\bigoplus}$ worth of embryos and planetesimals distributed in two areas, the first between 0.3 and 4AU (M$_{embryos}$ = 3.39 M$_{\bigoplus}$, M$_{planetesimals}$ = 3.44 M$_{\bigoplus}$) and a second mass distribution beyond the Jupiter-mass planet, located between 6 and 9AU (M$_{embryos}$ = 4.06 M$_{\bigoplus}$, M$_{planetesimals}$ = 1.30 M$_{\bigoplus}$). All solid bodies were distributed according to the relationships derived by \cite{kokubo:2000} regarding embryo mass, spacing and semi-major axis, assuming a separation of 5 to 10 mutual Hill radii in the inner region and 3 to 6 Hill radii in the outer region.

The surface mass density profile was taken to vary as r$^{-3/2}$ and was normalized to 10g cm$^{-2}$ at 1AU for the in-situ simulations and inner regions of the JD$_{5-1}$ simulations and normalized to 20g cm$^{-2}$ at 1AU for the outer region of the JD$_{5-1}$ simulations. An integration timestep of 0.5 days was utilized in the in-situ simulations. This imposes an artificial inner boundary on planetary formation as bodies located within $\sim$0.1AU from the host star are not accurately resolved (see numerical tests in Appendix A of \cite{raymond:2011}). Thus bodies located within this region are not considered in this study. A somewhat longer timestep of 2 days was applied for the JD$_{5-1}$ simulations, increasing the inner boundary to $\sim$0.2AU. All collisions were treated as perfect mergers (ie.~linear momentum is conserved and no mass is lost as fragments). Inward migration of the giant planet was simulated by including a drag force in the integrator. Migration in the JD$_{5-1}$ situations occurs over the first 10$^5$ years and the simulations were continued for a total of 50 Myr. The JD$_{5-1}$ simulations also incorporate the effects of gas drag on the planetesimal population. A MMSN \citep{hayashi:1981} is adopted, with the nebular density linearly decaying over a period of 10 Myr, and the planetesimals are given an effective diameter of 10 km for purposes of calculating the drag force. A detailed description of the implementation of the migration and the gas drag force is given in \cite{mandell:2007}. Four individual simulations were run for each of the scenarios considered with a random number generator determining unique initial distributions of solid material within the disk.

\textit{JD$_{5-0.25}$, JSD$_{5-0.25}$}: The JD$_{5-0.25}$ and JSD$_{5-0.25}$ simulations were run using the MERCURY hybrid symplectic integrator \citep{chambers:1999,chambers:1997}. As with the in-situ and JD$_{5-1}$ situations, both planetary embryos and a planetesimal swarm were included. Solid bodies were initially located between 0.25 and 4.5AU and between 6 and 9AU. Mass was again equally divided between the planetesimals and embryos in the inner region (M$_{embryos}$ = M$_{planetesimals}$ = 5 M$_{\bigoplus}$), while the outer region contained four times more mass in the embryos than the planetesimals (M$_{embryos}$ = 5.7 M$_{\bigoplus}$, M$_{planetesimals}$ = 1.3 M$_{\bigoplus}$), giving a total solid mass of 17 M$_{\bigoplus}$. As in the JD$_{5-1}$ simulations, the solid bodies were scaled according to the mass relation of \cite{kokubo:2000} and were randomly separated by 5 to 10 mutual Hill radii in the inner region and 3 to 6 Hill radii in the outer region.

The surface mass density profile was again taken to vary as r$^{-3/2}$, but was normalized to a marginally higher value of 13.2g cm$^{-2}$ at 1AU. The integration timestep was set to 2 days, which reduces the inner boundary for accurate integration and simulation to 0.05AU.  As with the JD$_{5-1}$ simulations, giant planet migration and gas drag on the planetesimals are incorporated into the integrator as described in \cite{mandell:2007}.  Migration of the giant planet again occurs over 10$^5$ years and the simulations were run for a total of 200 Myr. Five individual simulations were run for each migration scenario, however one of each was unusable.

\subsection{Chemical Simulations}
\begin{figure}
\plotone{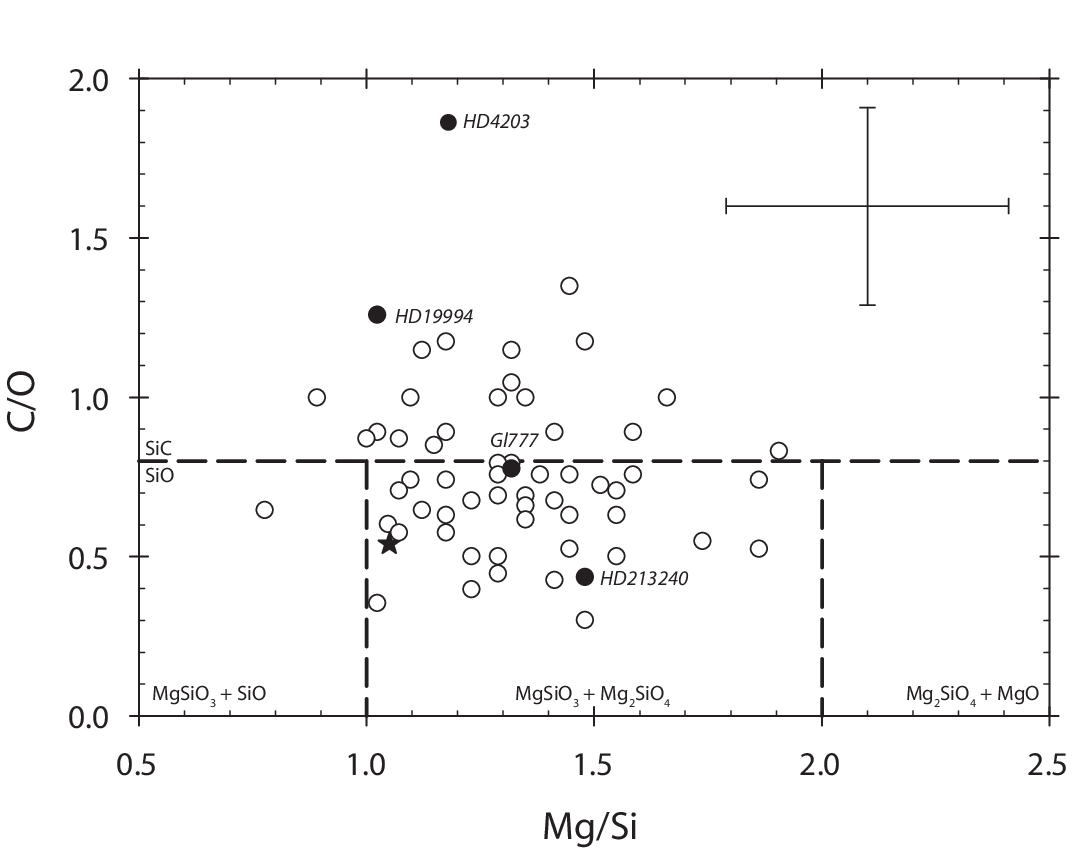}
\caption{Mg/Si vs. C/O for known planetary host stars with reliable stellar abundances.
Filled circles represent those systems selected for this study with their identifiers in italics. Stellar photospheric values were
taken from \cite{gilli}(Si, Mg), \cite{be}(Mg), \cite{ec}(C) and \cite{oxygen}(O). Solar values are shown by the black star and were taken
from \cite{asp}. The horizontal dashed line indicates a C/O value of 0.8 and marks the
transitions between a silicate-dominated composition and a carbide-dominated composition
at 10$^{-4}$ bar. The vertical dashed lines indicate the transition between the various Mg/Si regions. Dominant solid state composition for each region is shown. Average 2-$\sigma$ error bars for the observational estimates are shown in upper right.\label{regions}}
\end{figure}

In this study, we followed the same approach as in \cite{carter-bond:2012} and \cite{bond:2010a,bond:2010b}. Equilibrium between the solid material within the disk and the stellar nebula was assumed. As such, the final elemental composition of the simulated terrestrial planets can be determined by employing equilibrium condensation sequences. Following previous studies, HSC Chemistry (v. 5.1) was utilized to obtain condensation sequences for the 15 major solid forming elements (H, C, N, O, Na, Mg, Al, Si, P, S, Ca, Ti, Cr, Fe and Ni) via the Gibbs energy minimization method. While the dynamical simulations are generic and not designed to reflect any one specific system, stellar photospheric abundances for five known planetary host stars were selected for study - Solar, Gl777, HD4203, HD19994 and HD213240. These compositions were selected as they span the full range in both Mg/Si and C/O values observed for planetary host stars. Figure \ref{regions} displays the ratios for these systems, in addition to other known host stars for which Mg/Si and C/O values are available.

The potential range of host star compositions has recently come under some scrutiny as \cite{fortney:2012} illustrated the apparent disconnect between the composition of known planetary host stars (primarily F- and G-type stars) and lower mass stars of surveys such as the Sloan Digital Sky Survey. The high stellar C/O ratios observed for some planet host stars \citep{petigura:2011,delgado-mena:2010} are not observed within other large uniform samples \citep[e.g.][]{covey:2008}, leading to the claim that observed host star C/O values are systematically overestimated and calling into question the true nature of the underlying distribution in C/O values. Numerous possible causes for this disparity have been suggested but the issue remains unclear at this stage. As such, we will continue to utilize previously published stellar elemental abundances as the initial nebula composition until the issue is resolved. These still represent the best available way to estimate the composition of the primitive solids in the planetary systems and are ideal as the starting point for studies such as ours.  We will reexamine this assumption and repeat any simulations as necessary as new stellar abundances become available and as the issues surrounding the C/O distribution are resolved.

Observed stellar abundances for the four extrasolar planetary host stars were taken from \cite{oxygen, ec, gilli} and \cite{be}. These abundances were determined by the same research group from the same spectra in a uniform manner, thus minimizing any possible systematic errors between studies. Abundances of N and P have not been determined spectroscopically and were instead approximated based on the observed odd-even effect. The same method was applied for those stars without observed S values. All Solar values were taken from \cite{asp}. The input values used in HSC Chemistry (normalized to 10$^{6}$ Si atoms) for each system are shown in Table \ref{input_chem}. Equilibrium conditions at the disk midplane were obtained from radial pressure and temperature profiles adopted from \cite{hersant:2001} (see \cite{bond:2010b} for specific mass accretion rates utilized). Based on the best-fit to known Solar System planetary values \citep{bond:2010a}, disk conditions at an evolutionary time of t = 5$\times$10$^{5}$yr were utilized here. Further details of this method are given in \cite{bond:2010a,bond:2010b}.

It should be noted here that all of the solid-material compositions utilized in this study are based on chemical equilibrium between the solid material within the disk and the nebular gas. However, in the outer regions of the disk (beyond $\sim$ 6 AU), kinetic inhibition may prevent the reduction of CO to the equilibrium species CH$_{4}$ \citep{lewis:1980}. As such, equilibrium compositions may not best reflect the composition of solid species (including ices, clathrates and hydrates) forming in this region \citep{johnson:2012,wong:2008,gaidos:2000}. In order to maintain internal consistency, our models assume equilibrium-driven compositions throughout the entire disk. A separate study of non-equilibrium driven volatile compositions is currently underway to better address this issue.

\subsection{Combining Dynamics and Chemistry}
As in previous studies, combining the chemical and dynamical simulations was done by assuming that each body in the dynamical simulations retains the equilibrium composition of the region in which it is first located and contributes that same composition to the final planetary body.  Phase changes and outgassing by the bodies during migration are neglected. In keeping with the dynamical simulations, perfect collisions are assumed (i.e. no mass loss occurs during collision). Although we do determine the component bodies' initial mineralogical compositions, we only report final bulk planetary elemental abundance as we do not simulate planetary differentiation and evolution of the interior.

We limited our results to planets that underwent \textit{at least} two collisions. That is to say that we ignore those embryos that have simply survived for the simulation period and not accreted any additional material, primarily because of their extremely low masses.

\section{Results}
\subsection{Dynamical}
\begin{figure}
\plotone{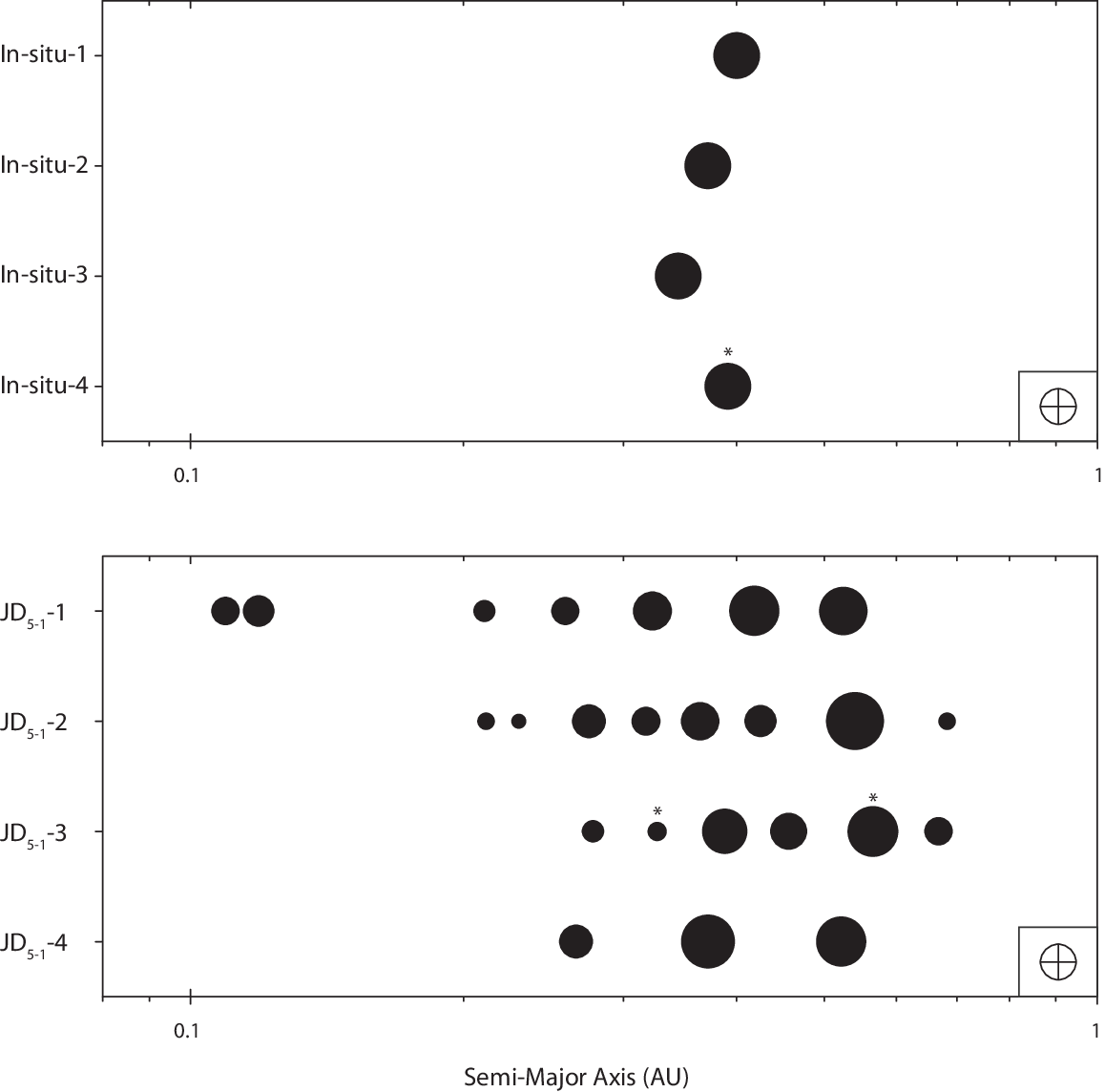}
\caption{Schematic of the results of the dynamical simulations for the in-situ scenario (top) and JD$_{5-1}$ scenario (bottom). Symbol size is proportional to the cube root of planetary mass. Earth is given in each panel for scale. The Jupiter-mass giant planet at 1 AU is omitted for clarity. Note that those planets with an * above them are shown in \ref{pie1} and \ref{pie2}.\label{system1}}
\end{figure}

\begin{figure}
\plotone{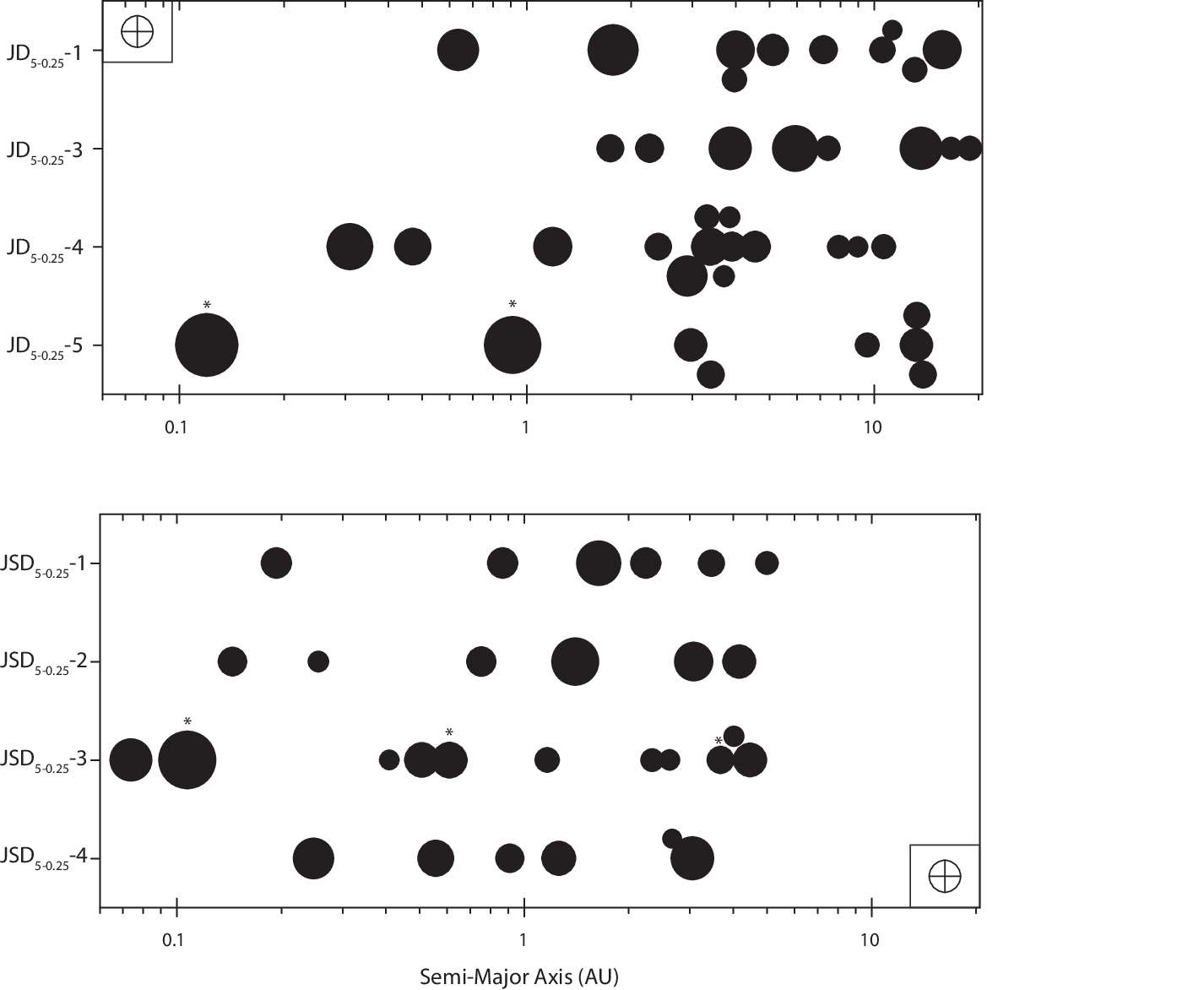}
\caption{Schematic of the results of the dynamical simulations for the JD$_{5-0.25}$ scenario (top) and JSD$_{5-0.25}$ scenario (bottom). Symbol size is proportional to the cube root of planetary mass. Elevation of some bodies above the plane is done only to resolve smaller, close-in bodies and is not an indication of planetary inclination. Earth is given in each panel for scale. The Jupiter-mass giant planet at 0.25 AU (both panels) and the Saturn-mass giant planet at 9.5 AU (bottom panel only) are omitted for clarity. Note that those planets with an * above them are shown in \ref{pie1} and \ref{pie2}.\label{system2}}
\end{figure}

The dynamics of terrestrial planet formation during giant planet migration has been examined in detail by numerous other studies \citep{fogg:2009,fogg:2007,fogg:2005,raymond:2006,zhou:2005} and is not the focus of this paper, so we provide only a basic discussion of the dynamical simulations. Terrestrial planets formed in all the simulations considered here. The architecture for each of the final systems is shown in Figures \ref{system1} and \ref{system2}. Note that these figures do not include any surviving planetesimals or embryos that did not undergo at least two collisions. Symbol radius scales as the cube root of planetary mass. Earth is shown in each panel for scale.

A single $\sim$ 1 M$_{\bigoplus}$ planet was produced for each of the in-situ simulations due to the limited mass available for terrestrial planet formation and the lack of damping via gas drag. For those simulations incorporating giant planet migration, multiple terrestrial planets were produced with masses ranging from 0.04 M$_{\bigoplus}$ to 4.15 M$_{\bigoplus}$. The planets produced by the JD$_{5-1}$ simulations have lower masses than those of the JD$_{5-0.25}$ and JSD$_{5-0.25}$ simulations. The vast majority of JD$_{5-1}$ terrestrial planets have masses less than 1 M$_{\bigoplus}$ with a maximum planetary mass of 2.25 M$_{\bigoplus}$. It is interesting to notice in Fig. 2 the pileup in mass just interior to the 2:1 mean motion resonance (at 0.62 AU), which is a direct consequence of shepherding of planetesimals and embryos during the giant planet's migration. Both the JD$_{5-0.25}$ and JSD$_{5-0.25}$ simulations have higher peak terrestrial planet masses (4.15 M$_{\bigoplus}$ and 3.23 M$_{\bigoplus}$, respectively) due to the higher degree of radial mixing and inward shepherding of material generated by the larger migration and possibly also because of the somewhat larger initial disk mass. Finally, the JD$_{5-1}$ simulations can be seen from Fig. \ref{system1} to not produce any terrestrial planets located beyond the final position of the gas giant, only stray embryos located beyond several AU. This mainly due to the longer accretion time required to produce a planet in that region and the relatively short (50 Myr) timescale of the simulations. For a more detailed discussion of the JD$_{5-0.25}$ and JSD$_{5-0.25}$ simulations, the reader is referred to \cite{mandell:2007}.

\subsection{Chemical}
The predicted bulk elemental abundance (in wt\%) for each of the simulated terrestrial planets is shown in Table \ref{results}. Giant planet migration during terrestrial planet formation can be seen to greatly alter the composition of the resultant terrestrial planets. When considering terrestrial planets that formed during giant planet migration, two broad types of terrestrial planet can be seen to be produced: those are composed of Mg-silicates and metallic Fe (in various amounts) (for disk compositions of Solar, HD213240 and the majority of Gl777) and those that are C-enriched, containing up to 47wt\% C in addition to Fe and Mg-silicate species (for disk composition of HD19994 and HD4203). Both types of planetary composition are similar to those produced by the in-situ simulations of the current study and those of \cite{bond:2010b}. However, giant planet migration is found to have two main consequences for the final composition of terrestrial planets: (1) The production of ``Earth-like" planets is increased, and (2) Hydrous phases are delivered to the dry inner regions of the system during the accretion process with a much greater efficiency. Each of these is discussed in turn below.

\subsubsection{Increased number of Earth-like planets}
\begin{figure}
\plotone{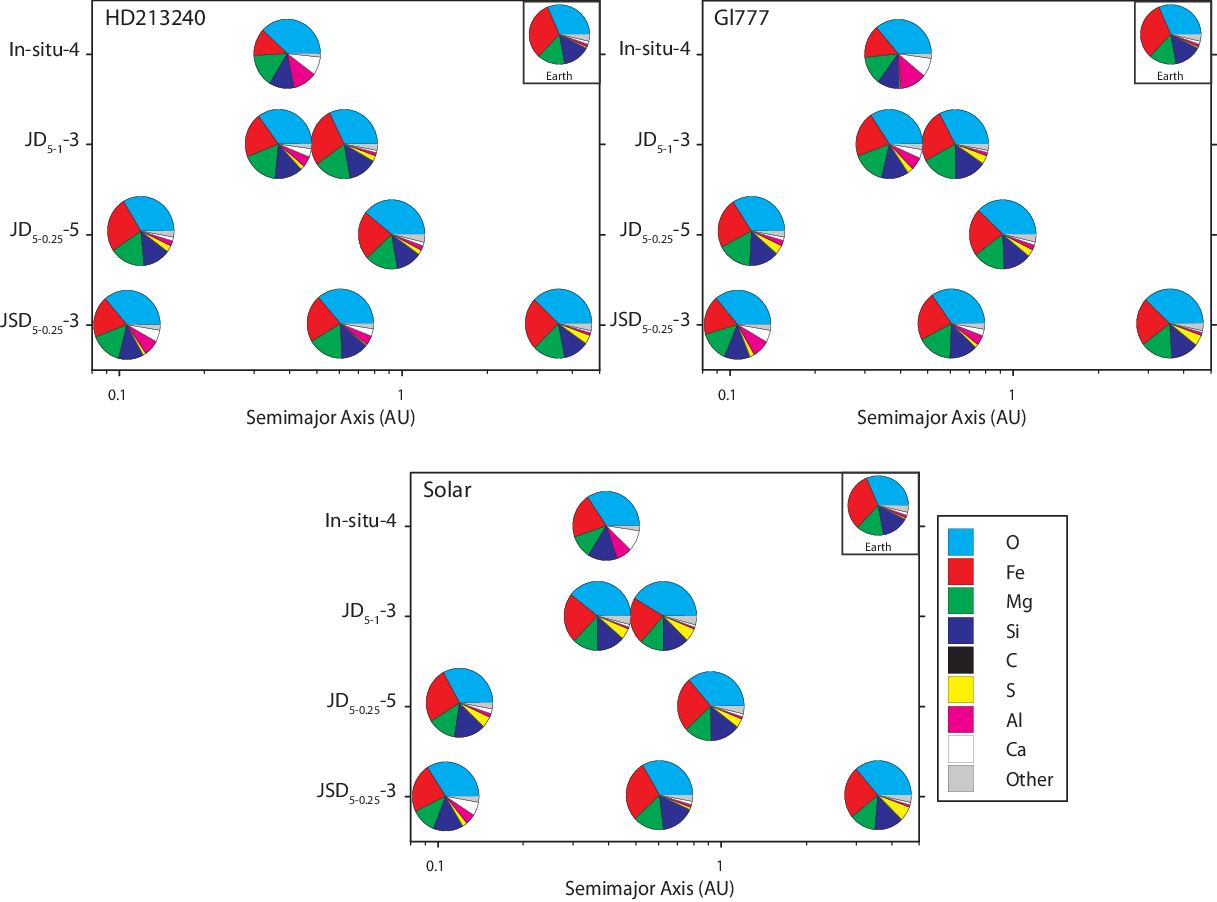}
\caption{Schematic of the bulk elemental planetary composition for planets produced assuming disk compositions with C/O $<$ 0.8. \emph{Upper Left:} Bulk elemental planetary compositions assuming a HD213240 composition disk. \emph{Upper Right:} Bulk elemental planetary compositions assuming a Gl777 composition disk. \emph{Bottom:} Bulk elemental planetary compositions assuming a Solar composition disk. All values are wt\% of the final simulated planet. For the sake of clarity, values are shown for a few indicative terrestrial planets produced in each of the four scenarios considered. Size of bodies is not to scale. Earth values taken from \cite{kargel:1993} are shown for comparison. Note that planets shown here were indicated with an * above them in \ref{system1} and \ref{system2}.\label{pie1}}
\end{figure}

\begin{figure}
\epsscale{.80}
\plotone{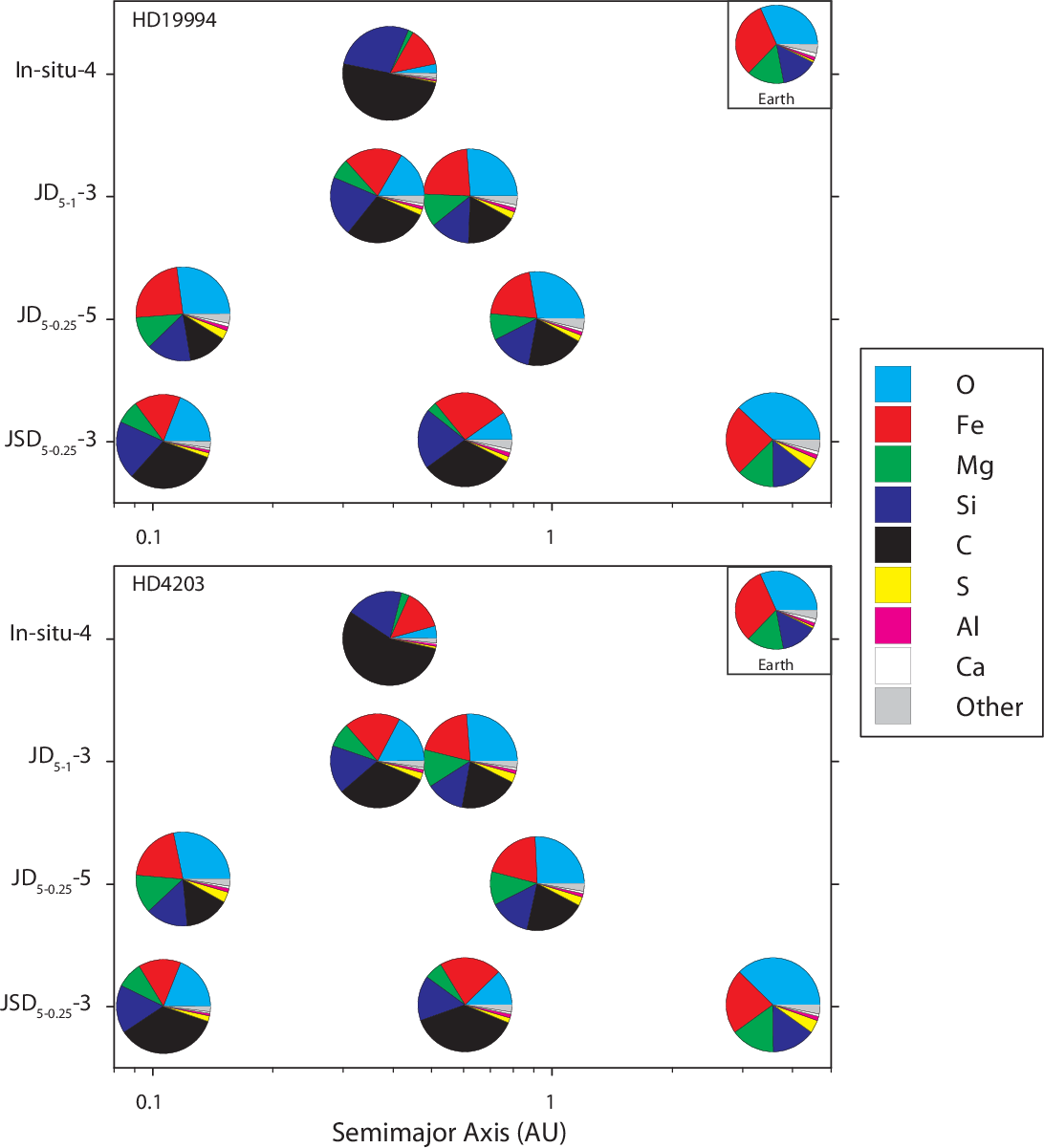}
\caption{Schematic of the bulk elemental planetary composition for planets produced assuming disk compositions with C/O $>$ 0.8. \emph{Top:} Bulk elemental planetary compositions assuming a HD19994 composition disk. \emph{Bottom:} Bulk elemental planetary compositions assuming a HD4203 composition disk. All values are wt\% of the final simulated planet. For the sake of clarity, values are shown for a few indicative terrestrial planets produced in each of the four scenarios considered. Size of bodies is not to scale. Earth values taken from \cite{kargel:1993} are shown for comparison. Note that planets shown here were indicated with an * above them in \ref{system1} and \ref{system2}.\label{pie2}}
\end{figure}

The most significant effect of giant planet migration on the compositions of simulated terrestrial planets is an increase in the relative abundance of Mg, Si, O and Fe, leading to an increased number of Earth-like planets. Figures \ref{pie1} and \ref{pie2} display representative bulk planetary abundances for terrestrial planets formed for all four migration scenarios for those disks with C/O$<$0.8 (Solar, HD213240 and Gl777)(Fig. \ref{pie1}) and those with C/O$>$0.8 (HD19994, HD4203)(Fig. \ref{pie2}). Throughout this paper, ``Earth-like'' does not imply an exact Earth twin in terms of bulk elemental abundances. Here we have followed \cite{bond:2010b} and use the term to refer to a planetary body that is primarily composed of metallic Fe and Mg-silicates with few other trace species present. Specifically, we have termed a planet to be Earth-like if the bulk elemental abundance for the major elements (O, Fe, Mg, Si) is within $\pm$25\% of the bulk Earth elemental abundances in \cite{kargel:1993}. By this definition, both Venus and Mars would also be considered to be Earth-like, thus the term should not be narrowly interpreted to mean that a planet is exactly like the Earth.

The increased abundance of Mg, Si, O and Fe (and consequent increased occurrence of Earth-like planets) reflects the composition of the solid material shepherded inward by the migrating giant planet. With giant planets on fixed orbits, terrestrial planets grow mainly by accreting objects that originated within $\sim 1$ AU of their final location (albeit with a tail that extends to larger distances; see \cite{raymond:2006b} for details). However, a migrating giant planet's strong dynamical influence on the small bodies in the disk (via shepherding and scattering) creates a large-scale mixing of solids with different initial compositions. The majority of solid material that is implanted into the planetary feeding zone originates from within 5AU from the host star. Assuming chemical equilibrium, for the majority of stars this material is dominated by Mg-silicates (pyroxene and olivine) and metallic Fe (between 0.5 and 4AU) and Mg-silicates and water ice beyond 4AU. Thus, the composition of solid material located in the inner terrestrial planet forming region of the disks (i.e within 0.5AU from the host star) changes from being primarily composed of refractory species (Ti, Al, Ca, O) to being dominated by Mg, Si, O and Fe. It should be noted, however, that the scattering induced by giant planet migration does not eliminate the C enrichment observed under equilibrium for those systems with C/O$>$0.8. C is located within $\sim$1.5AU from the host star (based on \cite{hersant:2001} disk profiles at 5$\times$10$^{5}$ years). Scattering via giant planet migration results in the introduction of Mg-silicate and metallic Fe material to this region, yielding a feeding zone composed of Mg, Si, O, Fe and C.

No Earth-like planets were produced in the in-situ simulations considered here (i.e. Jupiter-mass body stationary at 1AU). However, Earth-like planets were produced for Solar, HD213240 and Gl777 composition disks with a Jupiter-mass planet migrating from 5AU to 1AU (JD$_{5-1}$). These three systems have the lowest C/O values (0.54, 0.44 and 0.78, respectively). As such, the region of the disk interior to 5AU (and therefore the terrestrial planets' feeding zone) is C-free and dominated by silicate species, making the formation of an Earth-like planet more likely.

Earth-like planets were produced for all disk compositions considered, regardless of the C/O value, for the case of a Jupiter-mass planet migrating from 5AU to 0.25AU both with (JSD$_{5-0.25}$)and without (JD$_{5-0.25}$) a stationary Saturn-mass planet at 9.5AU. Earth-like planets are readily produced in systems which have undergone giant planet migration and may be common within extrasolar planetary systems. Such planets can from in any system, regardless of the C/O of the host star and therefore the system as a whole, if sufficient migration has occurred (high C/O systems require a higher extent of migration to produce an Earth-like planet). That is not to say that the C-rich planets of \cite{bond:2010b} are not produced. Instead, by incorporating giant planet migration we are now producing both Earth-like planets and C-rich planets within the same system (see JSD$_{5-0.25}$ simulations in both panels of Fig. \ref{pie2}).

Giant planet migration strongly affects the nature of the C-rich planets that form (see Table \ref{results} and Figure \ref{pie2}; see also \cite{bond:2010b}). With no giant planet migration, terrestrial planets composed of Fe, Si and more than 55wt\% C can form. However, including migration can significantly reduce the C enrichment, resulting in planets composed of C, Si, Fe and Mg, with C abundances varying from $<$1wt\% to 47wt\% and the majority containing $<$35wt\% C. This leads to a much more diverse range of C-enriched planets; many simulated planets (such as those produced with the disk composition of Gl777) could be considered C-enriched Earth-like planets as they contain a significant C abundance in conjunction with Earth-like abundances of Mg, Si, O and Fe.

The terrestrial planets in migration simulations that produced multiple terrestrial planets within $\sim$2.3 AU have a strongly homogenous bulk planetary composition, with planets located beyond $\sim$2 AU generally displaying a more Earth-like composition (namely lower C for the high-C/O systems and lower refractory components like Ca and Al for the lower-C/O systems). In contrast, \cite{bond:2010b} found the opposite trend in their simulations with static giant planets: strong radial planetary compositional gradients in multiple terrestrial planet systems due to narrowly constrained feeding zones, each with highly distinctive mineralogies. Indeed, the giant planet's migration causes such large-scale mixing that it effectively widens the feeding zone of each terrestrial planet to encompass the giant planet's entire pre- to post-migration orbital span, thus making the building blocks of each terrestrial planet virtually the same composition. It should also be noted that the close-in, refractory rich planets of \cite{bond:2010b} are not produced when giant planet migration is incorporated for the same reason, i.e. the refractory material in the inner disk is diluted by material from the outer regions, creating a more Earth-like composition.

\subsubsection{Delivery of hydrous phases}
In our simulations, giant planet migration systematically stimulates the delivery of large quantities of hydrous phases to the terrestrial planets (see also \cite{mandell:2007} and \cite{raymond:2006}). Here, the term ``hydrous phases'' refers to water ice and serpentine (Mg$_{3}$Si$_{2}$O$_{5}$(OH)$_{4}$), the aqueous alteration product of olivine.  Simulations are currently being finalized that contain a wider variety of volatile species, hydrates and clathrates and will be presented in a separate paper.

No water or serpentine was accreted by the simulated planets in any of the in-situ simulations. This is due to these planets' feeding zones being narrow and located within 1AU from the host star, combined with the fact that there was no material initially located beyond 1AU and therefore lacking any solid-state hydrous species within the simulation. Similar results were found in \cite{bond:2010a} and \cite{bond:2010b}. Including giant planet migration drastically increases the number of planets that accrete hydrous phases.  In the simulations with the largest degree of migration -- JSD$_{5-0.25}$ -- all of the simulated terrestrial planets obtained some amount of water during the accretion process.

\begin{figure}
\epsscale{.80}
\plotone{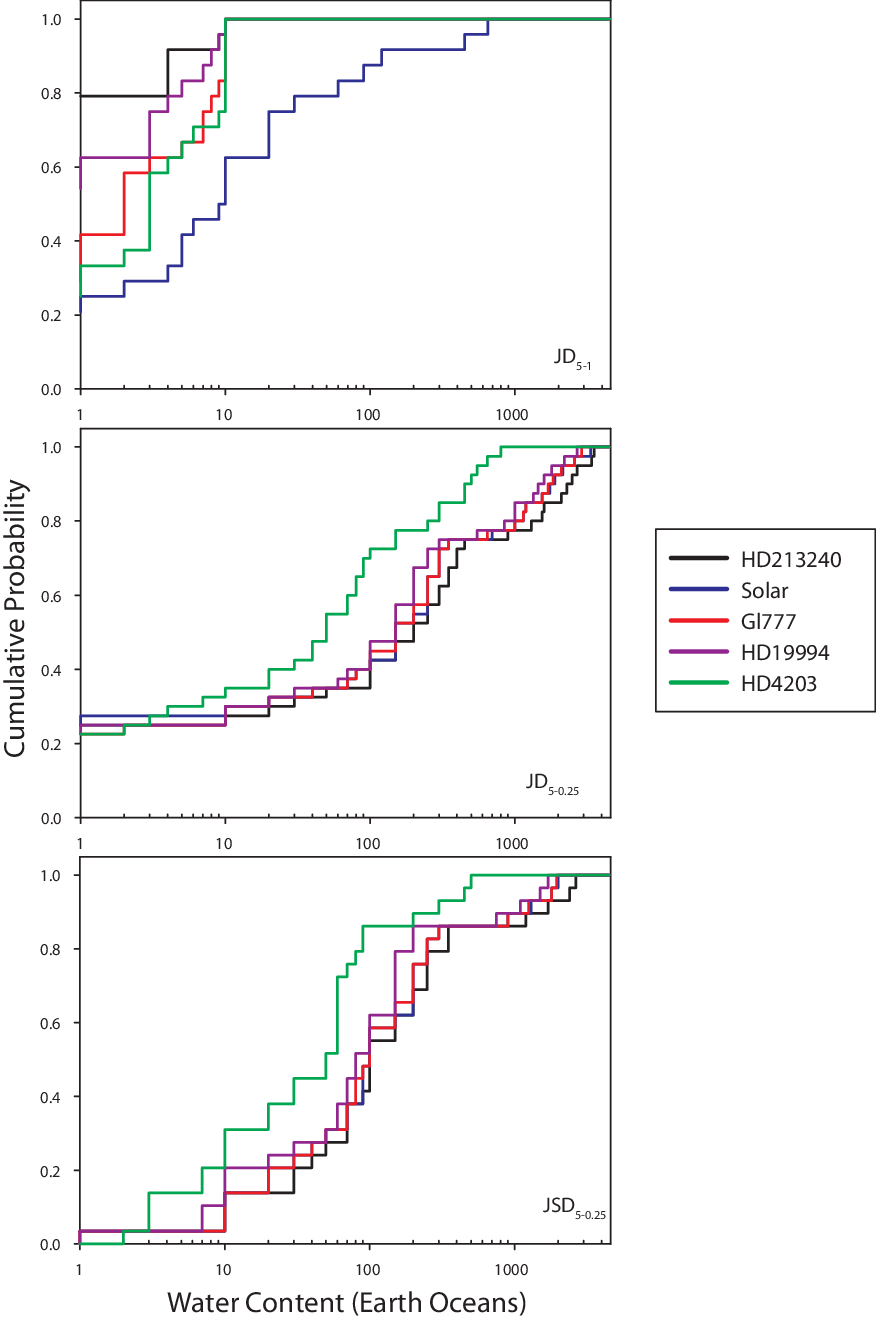}
\caption{Cumulative probability of planetary water content for the three migration scenarios considered. \emph{Top:} JD$_{5-1}$, \emph{Middle:} JD$_{5-0.25}$, \emph{Bottom:} JSD$_{5-0.25}$. Based on all four simulations run for each scenario and the values listed in Table \ref{oceans}.\label{probability}}
\end{figure}

The water abundance varies significantly with the disk composition, as increasing C/O values decreases the amount of water ice available \citep[see][for details]{gaidos:2000}. As such, HD213240 (C/O = 0.44) has the largest water reservoir available for direct accretion among our five disk compositions and HD4203 (C/O = 1.86) has the smallest. Assuming that all H accreted by the planet (whether as water ice or serpentine) is retained and converted to water, we obtain planetary water values of up to 4492 Earth oceans\footnote{1 Earth ocean mass of water = 1.4$\times$10$^{21}$kg = 2.34$\times$10$^{-4}$M$_{\bigoplus}$} for a HD213240 composition gas disk and up to 895 Earth oceans for a HD4203 composition gas disk. These results indicate that it is likely that the production of ``waterworld'' planets may result from giant planet migration, similar in nature to that of GJ1214b \citep{berta:2012}. Projected water contents for all compositions are listed in Table \ref{oceans} (sample only, complete table is available online) while Figure \ref{probability} displays the cumulative probability water abundance for each of the three migration scenarios and all five disk compositions considered. The water abundances reported here represent a significant increase on previous predictions. \cite{mandell:2007} estimated a planetary water content of $\sim$10\% water by mass, while \cite{raymond:2007} predicted planetary water contents ranging from 5 to 126 Earth oceans. Our water contents are higher as the simulations considered here include giant planet migration and its subsequent introduction of water-rich material to the terrestrial planetary feeding zone (\cite{raymond:2007} did not consider migration) and our water contents are based on equilibrium condensation sequences (both \cite{mandell:2007} and \cite{raymond:2007} base their water contents on observed values in chondritic meteorites).

Water delivery to close-in terrestrial planets occurs {\em during} giant planet migration via collisions of objects with a wide range of formation locations; water delivery to close-in planets after the end of migration is hindered by dynamical constraints \citep{mandell:2007}.  Water delivery to planets beyond a few tenths of an AU occurs via many impacts with volatile-rich bodies for the remainder of the simulation.  At this stage, many of the terrestrial planets have already accreted a significant fraction of their final mass, increasing the probability that the water (or at least a portion of it) will be retained and incorporated into the planet in some form. Having said that, the water content values listed in Table \ref{oceans} are likely to be upper limits as we do not consider water loss during collisions \citep{genda:2005,canup:2006} or via outgassing and hydrodyanmic escape from the planetesimals and embryos \citep{matsui:1986} occurring during the migration process.  It is unclear to exactly what extent these processes will impact the final amount of water delivered to a terrestrial planet, but a significant loss of volatile species is expected. For example, cometary observations have reported a sublimation rates reaching 10$^{30}$ molecules per second \citep{bockelee:2002}. Given the age of a planetary system, the loss from planetesimals and embryos before accretion via this mechanism alone can be severe. As such, it is expected that a significant reduction in global water content will occur. On the other hand, we have not considered other potential sources of water such as the oxidation of primitive H-rich atmospheres \citep{ikoma:2006} or adsorption of water vapor onto grains \citep{muralidharan:2011,drake:2005}. Both processes are likely to occur to some extent and offset H loss through sublimation and collisions. Additionally, although the planets are simulated to receive significant amounts of water, the vast majority of planets are located outside the traditional habitable zone. As such, it is unclear if such a body could retain an appreciable amount of water in the liquid form for a geologically and/or biologically significant period of time.

In the JD$_{5-0.25}$ simulations (and to a lesser extent JD$_{5-1}$), planetesimals and embryos initially located within 2 AU (and therefore lacking water) are often scattered out into the cooler, water-rich outer regions of the disk. Due to both the significantly longer timescale of accretion in the outer areas of the disk and the fact that our current approach does not accommodate compositional changes once condensation has occurred, the composition of these bodies continues to reflect that of the inner region of the system and they remain dry (i.e.contain no water) despite being located in regions of the disk well beyond 5AU where the composition of solid material is dominated by water ice and serpentine. It is unclear to what extent water ice may be present such a body, whether delivered after migration through collisional accretion, adsorption of water vapor from the disk or even compositional changes of the constituent minerals (such as serpentinization of olivine). Given their relatively small masses ($\lesssim$ 0.3M$_{\bigoplus}$) these scattered planets often remain on high-eccentricity orbits due to relatively weak damping from the disk \citep{tanaka:2004}.  We expect that these planets might simply continue to accrete in their new distant environments but on longer (Gyr) timescales than we have simulated.

Beyond biological considerations, water also significantly impacts planetary processes and properties. The impact of the predicted global water contents on these properties is deferred to Section \ref{int_water}.

\section{Observations, Implications and Discussion}
\subsection{White Dwarf Observations}
Direct comparison between the results of the current simulations and observations of extrasolar terrestrial planets is not yet feasible. Instead, we utilize polluted white dwarfs as an observational test of our results. The atmosphere of a white dwarf is composed of H and He only. Any element heavier than He and present in the atmosphere must have been accreted by the white dwarf during its mass loss phase from a planetary system orbiting the precursor star \citep{jura:2008,jura:2006,jura:2003}. See \cite{debes:2012} and \cite{bonsor:2011} for discussions of the details of planetary disruption and accretion.

Numerous observations \citep[e.g.][]{jura:2012a,jura:2012b,gansicke:2012,zuckerman:2011,klein:2011,klein:2010,zuckerman:2010,dufour:2010} have detected evidence of the accretion of solid material similar in composition to that of Earth and CI chondrites, the most primitive meteorites within the Solar System. For example, the white dwarf NLTT 43806 was observed to have Mg/Si values between 1.05 and 1.26, indicating the accretion of a solid body (or several bodies) with a bulk elemental composition similar to that of Earth \citep{zuckerman:2011}. Furthermore, \cite{jura:2012b} found that refractory elements in two polluted white dwarfs (GD 40 and G241-6) are also in agreement with bulk Earth elemental abundances. Few observations of C pollution in white dwarfs have been completed to date. Those that have reported C-abundances have all reported C values that are severely depleted compared to Solar \citep{desharnais:2008,jura:2006}. Combined together, these observations represent a growing body of evidence supporting the results of this study that ``Earth-like" planets appear to be common in extrasolar planetary systems. It also lends weight to both the decreased occurrence of extreme C-rich bodies suggested by this study and the claim by \cite{fortney:2012} that although C-dominated bodies may exist, they are likely to be rarer than previously estimated.

\subsection{Wet Earth Planetary Processes\label{int_water}}
Beyond its obvious biological implications, water significantly impacts planetary evolution.  Even if present on a terrestrial planet in a relatively small amount (as in our Solar System), water exerts a disproportionately large control on the evolution and internal dynamics of a given terrestrial planet. Numerous large scale planetary processes and properties such as mantle rheology, viscosity, differentiation, crystallization, volcanic eruptions and even atmospheric composition are strongly influenced by the bulk planetary water content \citep[e.g.][]{saal:2008,asimow:2003,hirth:1996}. Water also influences the interpretation of observations of transiting planets, as the water (and therefore H) content of a planet has been found to be a controlling factor for some mass-radius models of both solid and gaseous planets \citep{grasset:2009,adams:2008}. Given that giant planet migration can introduce significant quantities of water to a terrestrial planet during its formation process, we will briefly consider the possible impact that this may have on a terrestrial planet. Note that the discussion here is primarily focussed on water-rich Earth-like planets as we do not have the required data available to accurately describe the behavior of the more exotic C-planet compositions at high temperatures and pressures and with varying water contents. This also limits us from determining the equation of state for a C-rich planet and undertaking a detailed interior model similar to that done by \cite{valencia:2007a} for Si-dominated planets. The interior structures and processes of dry planets (i.e. those without water) remain unchanged from previous models and readers are referred to \cite{carter-bond:2012,delgado-mena:2010,bond:2010a,bond:2010b} for a detailed discussion.

The concept of a terrestrial planet with extremely high water content is not new \citep[e.g.][]{leger:2004,franck:2003,kuchner:2003}. If we assume that all (or at least the vast majority) of the water delivered via migration is retained by the planet, then we are likely to produce planets with internal structures similar to that proposed by \cite{leger:2004}---a metallic core overlain by a silicate mantle, dominated by olivine and pyroxene with significant surface water present. For planets located beyond $\sim$2.5 AU, this surface water is likely to manifest as a global ice layer. Such a planet may also have a global ocean layer (depending on planetary surface temperature and heat flux) and may also experience cryovolcanism and resurfacing (see \cite{fu:2010} for detailed simulations). The depth of the surface water (or ice, as the case may be) will directly depend on the amount of water delivered to and retained by the planet.

If, on the other hand, we assume that 97\% of the water delivered in the current simulations is lost (based on observed elemental depletions in ordinary chondrites with respect to the 50\% condensation temperature \citep{davis:2006}), then we are left with few surviving waterworlds, containing just 27 - 135 Earth-oceans of water (depending on disk composition). This assumption also brings us into line with the predicted water contents of \cite{raymond:2007}. Under this assumption, the majority of terrestrial planets contain less than 10 Earth-oceans of water, with a median water content of $\sim$5 Earth-oceans. We refer to this type of body as a wet Earth: a terrestrial planet containing $\lesssim$ 10 Earth-oceans of water, with the water present both on the surface of the planet and incorporated into the interior. One of the most significant impacts of such a scenario will be on plate tectonics and global convection. Water, especially planetary surface water, is absolutely crucial to plate tectonics on planets of all sizes \citep{regenauer-lieb:2001,oneill:2007,valencia:2007b}, with several studies finding that the presence of surface water, not a planets size, is the greatest determining factor in whether or not plate tectonics will occur on a given planet \citep{vanheck:2011,korenaga:2010}. Increasing water content significantly reduces the yield strength of mantle materials such as olivine \citep{asimow:2003}, thus decreasing the force needed to generate slippage along a pre-existing fault or plate boundary. Such slippage may lead to subduction, a driving force behind mobile lid convection \citep{moresi:1998}, as observed on Earth. An anhydrous crust, such as that of Venus, requires higher forces to create the same degree of slippage, making the occurrence of a stagnant lid in such a system increasing likely \citep{valencia:2007b}. The addition of water has the added effect of lowering the olivine solidus \citep{hirschmann:1999}, resulting in the production of small melt pockets deep within upwelling regions of the planetary mantle \citep{robinson:2001,mckenzie:1985}. The same solidus and stability effects will also result in the the production of melts comparable in composition to island-arc tholeiitic melts observed here on Earth with fractionation and segregation during ascent \citep[see][and references thereafter]{nicholls:1973}. The presence of significant amounts of water in the planetary interior may also increase the total amount of melt produced \citep{hirschmann:1999,asimow:2003} (as observed for oceanic floor basalts here on Earth) and marginally thicken the crust \citep{asimow:2003}. Extending these to simulated extrasolar planets, we can infer that due to high simulated water contents, plate tectonics is likely to be common on extrasolar terrestrial planets. Significantly lower mantle forces would be required to initiate and maintain a mobile lid convection regime, and melt production would be increased. Intraplate volcanism may initially occur on the planet, due to the increased depth of melting, but this would be temporary in nature and only last until the initial melt material was depleted (assuming the material is not refreshed via convection) \citep{kite:2009}.

\cite{elkins-tanton:2008} found that if an oxidizer such as water is present in sufficient quantities during differentiation, Fe particles $\lesssim$ 1cm in size are likely to oxidize before a planetary core can be produced. In the current simulations, water is delivered in almost all collisions occurring after $\sim$10$^{6}$ years, meaning that it would be present during the bulk of the differentiation process. Thus it is possible that the terrestrial planets simulated here may not contain a metallic core comparable to that of Earth, effectively prohibiting the production of a magnetic dynamo. It is not clear what effects such a structure would have on the planetary processes outlined above. Further work, beyond the scope of this study, is required on these issues.

\subsection{Planet Habitability}
Of the five disk compositions examined, three produced Earth-like planets within the habitable zone.  Although there are many factors that may influence planetary habitability, here we follow \cite{kasting:1993} and adopt 0.8 and 1.5AU as the inner and outer boundaries respectively of the habitable zone within the Solar System. Both the Solar composition disk and the Gl777 composition disk produced five Earth-like planets located within the habitable zone with migration of a Jupiter-mass body from 5AU to 0.25AU with a Saturn-mass body stationary at 9.5AU (JSD$_{5-0.25}$) and two Earth-like planets with migration of a Jupiter-mass body from 5AU to 0.25AU (JD$_{5-0.25}$). The planets range in mass from 0.056M$_{\bigoplus}$ to 3.055M$_{\bigoplus}$ and all but the lowest mass planets accreted significant amounts of water during the accretion process. Thus by the definition adopted here, we can state that these 14 simulated planets are potentially habitable.

Both of the disk compositions producing Earth-like planets within the habitable zone have C/O values below the 0.8 transition value. As such, we feel that despite the fact that systems with C/O values above 0.8 can produce Earth-like terrestrial planets, planetary systems orbiting a host star with C/O$<$0.8 continue to be the better candidates for future planet searches focussed on finding true Earth twins and undertaking detailed astrobiological studies. That is not to say that systems with C/O$>$0.8 cannot produce such a planet or host life. Until we can better understand the compositional variations that may occur within a disk, they simply are not the best selection for a large-scale survey.

\subsection{Biological Potential}
In order to host life, a system needs to not only contain water but also possess sufficient quantities of various critical elements. Based on our knowledge of life on Earth, these elements are H, C, N, O, P and S. Every single simulated planet in this study is deficient in several key biological elements. N is missing from all of planets. In addition, planets that formed in Solar- and HD213240- composition disks are depleted in C. One possible source for these elements may be that ices, clathrates and hydrates such as NH$_{3}$, CH$_{4}$ and CH$_{3}$OH are only produced in the outermost regions of the disk (assuming equilibrium) and as such need to be introduced to the inner regions of the disk. In a more realistic setting, this could occur either through cometary or asteroidal impacts or possibly through scattering of material as caused by giant planet migration. Simulations addressing this issue and incorporating numerous clathrates, hydrates and ices are currently being finalized.

\subsection{Model Sensitivity}
Before we conclude, a word of caution needs to be included. To study the chemical diversity of planets, there are three primary inherent variables that need to be constrained: the initial composition of the solids in the disk, the feeding zone of each planet, and the accretion time for each planet, meaning the timing (and consequences of) accretion events and impacts.  The approach taken here was not to explore the full range of possible values for each of these variables but rather to use chemical simulations to calculate a few different disk initial conditions, and dynamical simulations to calculate the accretion history for a limited set of planets.  Our approach has the advantage of being self-contained to within the assumptions of the models we have used, although we appreciate that other, perhaps drastically different values for these three variables may be possible. In the current simulations, the migration parameters have the greatest impact on the final planets produced. The extent to which the giant planet migrates greatly influences the final terrestrial planet composition. This is clearly seen when comparing the JD$_{5-1}$ and JD$_{5-0.25}$ simulations to the in-situ case examined here and the simulations of \cite{bond:2010b}. For high degrees of giant planet migration, increasingly homogenous terrestrial planet compositions are produced, while water and other volatile species may be present on many (if not all) planets. This is in stark contrast to the case of a limited (or non-existent) degree of migration (in-situ and JD$_{5-1}$ in this study). In this case, we are more likely to see planets similar in nature to those of \cite{bond:2010b} with extremely diverse compositions reflecting the condensation sequence within the disk. The exact timing of giant planet migration also effects terrestrial planet composition. If migration occurs after terrestrial planet accretion has largely completed, the terrestrial planet compositions will reflect the only the composition of the solid material immediately adjacent to their formation location (i.e. the results of \cite{bond:2010b}), resulting in a suite of planetary compositions that strongly reflect the condensation sequences within the disk. Migration may later alter the location of such a planet within the system but the composition should still reflect a specific portion of the condensation sequence. If, on the other hand, giant planet migration occurs before and/or during terrestrial planet formation (as considered here), the terrestrial planetary feeding zones will now be composed not only of material initially located immediately adjacent to the forming planet but also presumably sourced from a variety of chemical regions within the disk, allowing for less drastic compositional variations between planets within a given system.

We also only consider migration of the giant planet towards the host star. We do not allow for it to then move back further out in the planetary system, as hypothesized by \cite{walsh:2011a,veras:2004,masset:2001}. Similarly, the presence of other planetary bodies impacts final planetary composition (JD$_{5-0.25}$ vs JSD$_{5-0.25}$).  In addition, many of the known extra-solar giant planets have high eccentricities that are suggestive of violent dynamical histories such as planet-planet scattering \citep{raymond:2010,chatterjee:2008,juric:2008}), possibly triggered by migration \citep{libert:2011,moorhead:2005}.  The effect of such instabilities is in almost all cases detrimental to terrestrial planet formation; in extreme cases all terrestrial building blocks are driven into the central star \citep{raymond:2012,raymond:2011}.

On the chemical side, the evolutionary time selected for the disk conditions strongly influences the final terrestrial planet composition. The results displayed here are for the time of ``best fit" disk conditions as observed for the Solar System \citep{bond:2010a}. If condensation were to occur significantly later, the disk would be cooler, condensation fronts would have moved closer to the host star and bulk planetary composition may be altered. The biggest impact of such a scenario would be to increase the water and Mg-silicate content of a planet and potentially reduce the C content in C-enriched planets. Until we learn more about the timing of condensation of solids in the disk, the timing of planetesimal and embryo formation from those solids, and the compositional evolution throughout the lifetime of a disk, we are required to adopt those parameters that best fit the observable data we currently have at hand i.e. the terrestrial planets of the Solar System. Furthermore, the choice of disk model has the distinct potential to impact the final planetary abundances \citep{elser:2012}. The uncertainties introduced by specific disk models appears to be somewhat limited in scale and produces only relatively small changes in planetary composition. It does not appear to alter the general classification of a planet (i.e. an Earth-like planet produced with one radial P and T profile is still Earth-like when assuming a different model).

Considering the numerous assumptions required by studies of this nature and the effects they may have on the final planetary composition, it is worth restating that the results presented here are only intended to be indicative of the range of terrestrial planets possible within extrasolar planetary systems and to place better guidelines on future planet search and astrobiological activities. Systems that are believed to have unusual migrational and/or formation histories should be considered in detail individually.

\section{Conclusions}
We have modeled how the compositions of terrestrial planets were affected by four different scenarios for giant planet migration using a total of 16 high-resolution dynamical simulations (including 8 from \cite{mandell:2007}).  Using the simulated dynamics of accretion as input, we calculated bulk terrestrial planet compositions assuming disk compositions from five different stars known to host giant planets (Solar, Gl777, HD4203, HD19994 and HD213240).  This study is the first to consider in detail the impact that giant planet migration is expected to have on terrestrial planet compositions.

Giant planet migration has two key effects on the composition of the simulated terrestrial planets. First, migration increases the number of Earth-like planets produced by increasing the abundance of Mg-silicates and metallic Fe in the planets that form. For the largest degree of migration considered here, at least one Earth-like planet was produced in each simulation, regardless of the composition of the disk. This implies that Earth-like planets can be expected to be common within extrasolar planetary systems. It also indicates that the C-rich planetary bodies of \cite{carter-bond:2012} and \cite{bond:2010b} are end-member cases and that the true terrestrial planet composition in systems with bulk C/O $>$ 0.8 may be somewhat less extreme than previously considered. Second, giant planet migration makes possible the delivery of vast amounts of water (as ice and serpentine) to the terrestrial planets during the accretion process (see also Raymond et al 2006). This is likely to result in the production of waterworlds, if taken to the extreme case, or more likely produce wet Earths - Earth-like planets with an excess of water. Such planets would experience unique planetary processes, given the strong effects of water on the nature of planetary processing.

Numerous Earth-like habitable planets are produced in our simulations, primarily for disk compositions with C/O $<$ 0.8. Despite the fact that high C/O systems may produce Earth-like planets, systems with bulk compositions closer to Solar remain our best hope for finding a truly habitable Earth twin. All terrestrial planets modeled here are found to be lacking key biological elements, but more advanced models that include a broader selection of volatile species (i.e. ices, clathrates and hydrates) are expected to result in the delivery of the required elements, producing of terrestrial planets favorable to the development of life.

\acknowledgments
J. C. Carter-Bond and D. P. O'Brien were funded by grant number NNX09AB91G from NASA's Origins of Solar Systems Program and grant NNX10AH49G from NASA's Fellowship for Early Career Researchers Program. J. C. Carter-Bond was also supported under Australian Research Council's Discovery Projects funding scheme (project number DP110104526). S.N.R. is grateful to the CNRS's PNP program and to the NASA Astrobiology Institute's Virtual Planetary Laboratory lead team. We would like to thank the anonymous reviewer for their comments and suggestions.
\appendix
\section{Online Material: Tables}
\clearpage
\begin{deluxetable}{cccccccccccccccccc}
\tabletypesize{\footnotesize}
\tablecolumns{18}
\tablewidth{0pt}
\rotate
\tablecaption{Predicted bulk elemental abundances for all simulated extrasolar terrestrial planets. All values are in wt\% of the final predicted planet for all five disk compositions examined. Planet number increases with increasing distance of the final planet from the host star. Final mass and orbital semi-major axis are listed.}
\tablehead{\colhead{Simulation}           & \colhead{M}& \colhead{a}& \colhead{H}           & \colhead{Mg}      &
\colhead{O}          & \colhead{S}  &
\colhead{Fe}          & \colhead{Al}    &
\colhead{Ca}  & \colhead{Na}  &
\colhead{Ni} & \colhead{Cr}& \colhead{P}  &
\colhead{Ti}          & \colhead{Si}    &\colhead{N}    &
\colhead{C}\\
& \colhead{(M$_{\bigoplus}$})& \colhead{(AU)}& & &
& && &
& && && & &}
\startdata
\multicolumn{18}{c}{Solar Composition Disk}\\ \hline																															
In-situ-1-3	&	1.260	&	0.400	&	0.00	&	11.31	&	34.23	&	0.00	&	21.44	&	7.11	&	9.19	&	0.06	&	1.48	&	0.26	&	0.08	&	0.43	&	14.42	&	0.00	&	0.00	 \\
	&		&		&		&		&		&		&		&		&		&		&		&		&		&		&		&		&		 \\
In-situ-2-3	&	1.245	&	0.372	&	0.00	&	11.02	&	34.44	&	0.00	&	20.74	&	7.54	&	9.75	&	0.08	&	1.42	&	0.25	&	0.08	&	0.45	&	14.22	&	0.00	&	0.00	 \\
	&		&		&		&		&		&		&		&		&		&		&		&		&		&		&		&		&		 \\
In-situ-3-3	&	1.100	&	0.345	&	0.00	&	10.79	&	34.69	&	0.00	&	20.10	&	7.86	&	10.16	&	0.05	&	1.37	&	0.25	&	0.08	&	0.47	&	14.17	&	0.00	&	0.00	 \\
	&		&		&		&		&		&		&		&		&		&		&		&		&		&		&		&		&		 \\
In-situ-4-3	&	1.178	&	0.391	&	0.00	&	10.97	&	34.39	&	0.00	&	20.80	&	7.57	&	9.78	&	0.07	&	1.46	&	0.25	&	0.08	&	0.45	&	14.18	&	0.00	&	0.00	 \\
\enddata																															
\end{deluxetable} 
\clearpage
\begin{deluxetable}{cccccc}
\tabletypesize{\footnotesize}
\tablecolumns{6}
\tablewidth{0pt}
\tablecaption{Predicted water content for all simulated extrasolar terrestrial planets. All values are in Earth oceans for all five disk compositions examined. Planet number increases with increasing distance of the initial embryo from the host star. Disk compositions are shown in order of increasing C/O value.}
\tablehead{\colhead{Simulation}           &
\multicolumn{5}{c}{Disk Composition}\\
& \colhead{HD213240}& \colhead{Solar}& \colhead{Gl777}           & \colhead{HD19994}      &
\colhead{HD4203}}
\startdata
In-situ-1-3	&	0.00	&	0.00	&	0.00	&	0.00	&	0.00	 \\
	&		&		&		&		&		 \\
In-situ-2-3	&	0.00	&	0.00	&	0.00	&	0.00	&	0.00	 \\
	&		&		&		&		&		 \\
In-situ-3-3	&	0.00	&	0.00	&	0.00	&	0.00	&	0.00	 \\
	&		&		&		&		&		 \\
In-situ-4-3	&	0.00	&	0.00	&	0.00	&	0.00	&	0.00	 \\
	&		&		&		&		&		 \\
\enddata																															
\end{deluxetable}

\clearpage

\clearpage

\clearpage

\clearpage

\clearpage

\clearpage

\clearpage

\begin{deluxetable}{crrrrr}
\tabletypesize{\footnotesize} \tablecolumns{11} \tablewidth{0pt} \tablecaption{Target star elemental abundances as number of atoms and
normalized to 10$^{6}$Si atoms. See text for references.\label{input_chem}}
\tablehead{\colhead{Element}           &\colhead{HD213240}           & \colhead{Solar} & \colhead{Gl777}      &
\colhead{HD19994}          & \colhead{HD4203}}
\startdata
Fe	&	1.00	$\times$10$^{6}$	&	8.71	$\times$10$^{5}$	&	8.32	$\times$10$^{5}$	&	8.51	$\times$10$^{5}$	&	7.59	$\times$10$^{5}$	\\
C	&	1.23	$\times$10$^{7}$	&	7.59	$\times$10$^{6}$	&	1.15	$\times$10$^{7}$	&	1.48	$\times$10$^{7}$	&	1.05	$\times$10$^{7}$	\\
O	&	2.82	$\times$10$^{7}$	&	1.41	$\times$10$^{7}$	&	1.48	$\times$10$^{7}$	&	1.17	$\times$10$^{7}$	&	5.62	$\times$10$^{6}$	\\
Na	&	8.13	$\times$10$^{4}$	&	4.57	$\times$10$^{4}$	&	6.31	$\times$10$^{4}$	&	1.07	$\times$10$^{5}$	&	5.75	$\times$10$^{4}$	\\
Mg	&	1.48	$\times$10$^{6}$	&	1.05	$\times$10$^{6}$	&	1.32	$\times$10$^{6}$	&	1.02	$\times$10$^{6}$	&	1.17	$\times$10$^{6}$	\\
Al	&	1.07	$\times$10$^{5}$	&	7.24	$\times$10$^{4}$	&	1.05	$\times$10$^{5}$	&	1.02	$\times$10$^{5}$	&	9.77	$\times$10$^{4}$	\\
Si	&	1.00	$\times$10$^{6}$	&	1.00	$\times$10$^{6}$	&	1.00	$\times$10$^{6}$	&	1.00	$\times$10$^{6}$	&	1.00	$\times$10$^{6}$	\\
S	&	2.95	$\times$10$^{5}$	&	4.27	$\times$10$^{5}$	&	3.31	$\times$10$^{5}$	&	2.40	$\times$10$^{5}$	&	2.63	$\times$10$^{5}$	\\
Ca	&	5.75	$\times$10$^{4}$	&	6.31	$\times$10$^{4}$	&	4.79	$\times$10$^{4}$	&	5.62	$\times$10$^{4}$	&	4.07	$\times$10$^{4}$	\\
Ti	&	2.95	$\times$10$^{3}$	&	2.45	$\times$10$^{3}$	&	3.31	$\times$10$^{3}$	&	2.45	$\times$10$^{3}$	&	2.57	$\times$10$^{3}$	\\
Cr	&	1.20	$\times$10$^{4}$	&	1.35	$\times$10$^{4}$	&	1.12	$\times$10$^{4}$	&	1.23	$\times$10$^{4}$	&	1.02	$\times$10$^{4}$	\\
Ni	&	5.50	$\times$10$^{4}$	&	5.25	$\times$10$^{4}$	&	5.13	$\times$10$^{4}$	&	5.50	$\times$10$^{4}$	&	4.79	$\times$10$^{4}$	\\
	&			&			&			&			&			 \\
C/O	&	\multicolumn{1}{c}{0.44}		&	\multicolumn{1}{c}{0.54}		&	\multicolumn{1}{c}{0.78}		&	\multicolumn{1}{c}{1.26}		&	\multicolumn{1}{c}{1.85}\\
Mg/Si	&	\multicolumn{1}{c}{1.48}		&	\multicolumn{1}{c}{1.05}		&	\multicolumn{1}{c}{1.32}		&	\multicolumn{1}{c}{1.02}		&	\multicolumn{1}{c}{1.18}\\
\enddata
\end{deluxetable}

\clearpage
\begin{deluxetable}{cccccccccccccccccc}
\tabletypesize{\footnotesize}
\tablecolumns{18}
\tablewidth{0pt}
\rotate
\tablecaption{Predicted bulk elemental abundances for all simulated extrasolar terrestrial planets. All values are in wt\% of the final predicted planet for all five disk compositions examined. Planet number increases with increasing distance of the initial embryo from the host star. Final mass and orbital semi-major axis are listed.\label{results}}
\tablehead{\colhead{Simulation}           & \colhead{M}& \colhead{a}& \colhead{H}           & \colhead{Mg}      &
\colhead{O}          & \colhead{S}  &
\colhead{Fe}          & \colhead{Al}    &
\colhead{Ca}  & \colhead{Na}  &
\colhead{Ni} & \colhead{Cr}& \colhead{P}  &
\colhead{Ti}          & \colhead{Si}    &\colhead{N}    &
\colhead{C}\\
& \colhead{(M$_{\bigoplus}$})& \colhead{(AU)}& & &
& && &
& && && & &}
\startdata
\multicolumn{18}{c}{Solar Composition Disk}\\ \hline																															
In-situ-1-3	&	1.260	&	0.400	&	0.00	&	11.31	&	34.23	&	0.00	&	21.44	&	7.11	&	9.19	&	0.06	&	1.48	&	0.26	&	0.08	&	0.43	&	14.42	&	0.00	&	0.00	 \\	
	&		&		&		&		&		&		&		&		&		&		&		&		&		&		&		&		&		 \\	
In-situ-2-3	&	1.245	&	0.372	&	0.00	&	11.02	&	34.44	&	0.00	&	20.74	&	7.54	&	9.75	&	0.08	&	1.42	&	0.25	&	0.08	&	0.45	&	14.22	&	0.00	&	0.00	 \\	
	&		&		&		&		&		&		&		&		&		&		&		&		&		&		&		&		&		 \\	
In-situ-3-3	&	1.100	&	0.345	&	0.00	&	10.79	&	34.69	&	0.00	&	20.10	&	7.86	&	10.16	&	0.05	&	1.37	&	0.25	&	0.08	&	0.47	&	14.17	&	0.00	&	0.00	 \\	
	&		&		&		&		&		&		&		&		&		&		&		&		&		&		&		&		&		 \\	
In-situ-4-3	&	1.178	&	0.391	&	0.00	&	10.97	&	34.39	&	0.00	&	20.80	&	7.57	&	9.78	&	0.07	&	1.46	&	0.25	&	0.08	&	0.45	&	14.18	&	0.00	&	0.00	 \\	
	&		&		&		&		&		&		&		&		&		&		&		&		&		&		&		&		&		 \\	
\enddata																															
\tablecomments{Table \ref{results} is published in its entirety in the
electronic edition of the {\it Astrophysical Journal}.  A portion is
shown here for guidance regarding its form and content.}
\end{deluxetable} 
\clearpage
\begin{deluxetable}{cccccc}
\tabletypesize{\footnotesize}
\tablecolumns{6}
\tablewidth{0pt}
\tablecaption{Predicted water content for all simulated extrasolar terrestrial planets. All values are in Earth oceans for all five disk compositions examined. Planet number increases with increasing distance of the initial embryo from the host star. Disk compositions are shown in order of increasing C/O value.\label{oceans}}
\tablehead{\colhead{Simulation}           &
\multicolumn{5}{c}{Disk Composition}\\
& \colhead{HD213240}& \colhead{Solar}& \colhead{Gl777}           & \colhead{HD19994}      &
\colhead{HD4203}}
\startdata
In-situ-1-3	&	0.00	&	0.00	&	0.00	&	0.00	&	0.00	 \\
	&		&		&		&		&		 \\
In-situ-2-3	&	0.00	&	0.00	&	0.00	&	0.00	&	0.00	 \\
	&		&		&		&		&		 \\
In-situ-3-3	&	0.00	&	0.00	&	0.00	&	0.00	&	0.00	 \\
	&		&		&		&		&		 \\
In-situ-4-3	&	0.00	&	0.00	&	0.00	&	0.00	&	0.00	 \\
	&		&		&		&		&		 \\
JD$_{5-1}$-1-3	&	0.00	&	2.62	&	1.76	&	0.66	&	2.02	 \\
JD$_{5-1}$-1-4	&	4.79	&	10.03	&	7.79	&	5.33	&	5.52	 \\
JD$_{5-1}$-1-5	&	0.00	&	0.64	&	0.30	&	0.00	&	0.00	 \\
JD$_{5-1}$-1-6	&	0.00	&	0.63	&	0.00	&	0.00	&	0.00	 \\
JD$_{5-1}$-1-7	&	0.00	&	5.25	&	2.64	&	1.62	&	4.04	 \\
JD$_{5-1}$-1-8	&	0.00	&	14.37	&	9.32	&	4.88	&	10.63	 \\
JD$_{5-1}$-1-9	&	14.35	&	22.79	&	18.24	&	12.08	&	12.42	 \\
	&		&		&		&		&		 \\
JD$_{5-0.25}$-1-3	&	474.95	&	369.39	&	363.00	&	311.10	&	107.31	 \\
JD$_{5-0.25}$-1-4	&	1348.42	&	1021.62	&	1009.62	&	865.98	&	278.88	 \\
JD$_{5-0.25}$-1-5	&	0.00	&	0.00	&	0.00	&	0.00	&	0.00	 \\
JD$_{5-0.25}$-1-6	&	2533.93	&	1906.04	&	1881.03	&	1616.43	&	505.01	 \\
JD$_{5-0.25}$-1-7	&	928.36	&	700.13	&	691.14	&	593.57	&	187.13	 \\
JD$_{5-0.25}$-1-8	&	277.15	&	208.47	&	205.74	&	176.80	&	55.24	 \\
JD$_{5-0.25}$-1-9	&	0.00	&	0.00	&	0.00	&	0.00	&	0.00	 \\
JD$_{5-0.25}$-1-10	&	0.00	&	0.00	&	0.00	&	0.00	&	0.00	 \\
JD$_{5-0.25}$-1-11	&	0.00	&	0.00	&	0.00	&	0.00	&	0.00	 \\
JD$_{5-0.25}$-1-12	&	2108.80	&	1589.87	&	1579.08	&	1355.70	&	463.26	 \\

\enddata	
\tablecomments{Table \ref{oceans} is published in its entirety in the
electronic edition of the {\it Astrophysical Journal}.  A portion is
shown here for guidance regarding its form and content.}																														
\end{deluxetable} 
\clearpage
\end{document}